# Enhancing Cloud-Based Large Language Model Processing with Elasticsearch and Transformer Models

Chunhe Ni[1]*Jiang Wu[1],Hongbo Wang[2],Wenran Lu[3],Chenwei Zhang[4]

[1]*Computer Science,University of Texas at Dallas, Richardson, TX, USA
[1]Computer Science,University of Southern California, Los Angeles, CA, USA
[2]Computer Science, University of Southern California, Los Angeles, CA,USA
[3]Electrical Engineering, University of Texas at Austin, Austin, TX,USA
[4]Electrical and Computer Engineering,University of lllinois Urbana-Champaign,Urbana, IL,USA
* Corresponding author: nichunhe@outlook.com

## ABSTRACT

Large Language Models (LLMs) are a class of generative AI models built using the Transformer network, capable of leveraging vast datasets to identify, summarize, translate, predict, and generate language. LLMs promise to revolutionize society, yet training these foundational models poses immense challenges. Semantic/vector search within large language models is a potent technique that can significantly enhance search result accuracy and relevance. Unlike traditional keyword-based search methods, semantic search utilizes the meaning and context of words to grasp the intent behind queries and deliver more precise outcomes. Elasticsearch emerges as one of the most popular tools for implementing semantic search—an exceptionally scalable and robust search engine designed for indexing and searching extensive datasets. In this article, we delve into the fundamentals of semantic search and explore how to harness Elasticsearch and Transformer models to bolster large language model processing paradigms. We gain a comprehensive understanding of semantic search principles and acquire practical skills for implementing semantic search in real-world model application scenarios.

## 1. INTRODUCTION

Recent years have witnessed the introduction of a series of large-scale language models (Brown et al., 2020; Lieber et al., 2021; Rae et al., 2021; Smith et al., 2022; Thopilan et al., 2022), with the current largest dense language model boasting over 500 billion parameters. These large autoregressive Transformer models (Vaswani et al., 2017) have exhibited impressive performance across various evaluation protocols, including zero-shot learning, few-shot learning, and fine-tuning.Large Language Models (LLMs) have brought about a revolution in natural language processing, showcasing remarkable advancements in translation, summarization, and language generation tasks. However, deploying LLMs in cloud-based environments presents significant challenges in scalability and efficiency. To tackle these challenges, technologies such as Elasticsearch and Transformer models have emerged as pivotal components. While we arrive at a similar conclusion, we estimate that training large models should be significantly more extensive than suggested by the authors. Specifically, given a tenfold increase in compute budget, they propose that model size should increase by 5.5 times, while the number of training tokens should only increase by 1.8 times. In contrast, we find that model size and the number of training tokens should scale in equal proportion.By reassessing the logic, we underscore the importance of optimizing computational resources for training large language models while considering both model size and training token count in tandem.

Elasticsearch is a highly scalable search engine, optimized for indexing and querying large datasets, while Transformer models, exemplified by architectures like BERT and GPT, excel at capturing intricate linguistic patterns. This paper aims to explore how Elasticsearch can enhance the processing capabilities of LLMs, particularly Transformer models, in cloud-based environments.By highlighting the significance of LLMs, discussing cloud-based processing challenges, and introducing Elasticsearch and Transformer models, this paper lays the groundwork for an in-depth exploration of their integration to improve language processing efficiency in cloud environments.

## 2. RELATED WORK

### 2.1 Large language models and their applications

In the rapidly evolving field of artificial intelligence (AI) and machine learning (ML) models, Large language models (LLMs) have become part of the innovative evolution of the way we interact with technology and information. Although the technology is still in its infancy, large language models have become an important part of critical business in numerous industries around the world. Large language models are generative AI models that understand the inputs to human language and generate the corresponding outputs. Models like ChatGPT are trained using large amounts of data to detect patterns in human natural language and predict the probability of the next word based on context. Conversational AI chatbots like GPT-3 respond to prompts through text generation. This predictive power enables LLMs to generate sentences that are broadly coherent and context-relevant on a particular topic.

Large Language models (LLMs) use two main principles: machine learning and natural language processing (NLP). LLM training is exposed to large amounts of textual data, allowing it to learn nuances of language and context through predictions. Unlike traditional language-based models and algorithms, LLMs does not have pre-programmed language syntax rules. Instead, it learns itself through exposure to language, similar to how humans learn their native language. The transformer model assigns probabilities to all possible subsequent words, choosing only the one with the highest probability based on the data set, encoder, and learning model it is trained on. Zero-shot zero-time learning involves the model recognizing and responding to data that it has not processed in training, although this ability is still less common and often inaccurate.

### 2.2 Overview of Elasticsearch

Elasticsearch is a powerful and extensible open source search engine based on Lucene libraries. It is designed to process large amounts of unstructured data and provide fast and accurate search results. Elasticsearch uses a distributed architecture, which means it can scale horizontally across multiple servers to handle large amounts of data and traffic. Elasticsearch is built on RESTful apis, which makes it easy to integrate with a variety of programming languages and tools. It supports complex search queries, including full-text search, faceted search, and geographic search. Elasticsearch also provides a powerful aggregation framework that allows you to perform complex data analysis on search results.

**How Elasticsearch alleviates LLM problems:**

Provide data context and integrate with ChatGPT or other LLMS: Elasticsearch can store and manage large amounts of data and provide rich contextual information that helps LLMS understand query intent and produce more accurate results.

Support for built-in models (any third-party model) : Elasticsearch can access a variety of pre-trained language models, including ChatGPT and other LLMS, to provide users with more flexible options.

• Built-in Elastic Learned Sparse Encoder model: This model can efficiently represent text vectorically for easy vector search and analysis.

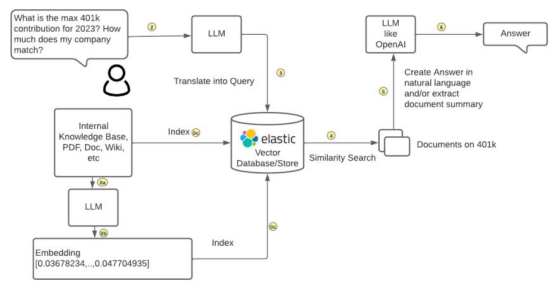

**Figure 1.** Elasticsearch implements process architecture

**Advantages of Elasticsearch as a vector database:**

Efficient hybrid search: Elasticsearch can perform both text search and vector search at the same time, satisfying a variety of application scenarios.

Massive data storage: Elasticsearch can store and manage large amounts of text and vector data, providing rich data resources for LLMS.High performance queries: Elasticsearch queries are very fast and can meet the needs of real-time retrieval.

Elasticsearch and LLM combine in three ways:

Mode 1: Elasticsearch and LLM Use Elasticsearch as a vector store and integrate with LLM.If the paper does not have the margins shown in Table 1, it will not upload properly.

## 2.3 Process the Transformer model and large language model

The Transformer architecture is the basic building block for all Language Models with Transformers (LLM). The Transformer architecture was introduced in the paper "Attention is all you need" published in December 2017. A simplified version of the Transformer architecture is as follows:

There are seven important components in the transformer architecture. Let's go through each of these components and see what they do in a simplified way:

1. Input and input embedding: The tags entered by the user are treated as inputs to the machine learning model. However, the model can only understand numbers, not text, so these inputs need to be converted into a numeric format called "input embedding." Input embeddings represent words as numbers, which can then be processed by a machine learning model. These embeddings act like a dictionary, helping the model understand the meaning of words by placing them in mathematical Spaces where similar words are close to each other. During training, the model learns how to create these embeddings so that similar vectors represent words with similar meanings.

2. Positional coding: In natural language processing, the order of words in a sentence is crucial to determining the meaning of the sentence. However, traditional machine learning models, such as neural networks, do not inherently understand the order of inputs. To address this challenge, positional coding can be used to encode the position of each word in the input sequence as a set of numbers. These numbers can be entered into the Transformer model along with the input embeddings. By incorporating positional coding into the Transformer architecture, GPT can more effectively understand the order of words in a sentence and generate syntactically correct and semantically meaningful output.

## 2.4 Related work in using Elasticsearch with Transformer models

Related work in using Elasticsearch with Transformer models or similar technologies has gained significant attention in recent years due to the growing need for efficient and scalable solutions in natural language processing (NLP). Here, we provide a review of the key studies and developments in this domain:

Transformer Models for Information Retrieval: Transformer models, such as BERT (Devlin et al., 2018) and T5 (Raffel et al., 2020), have shown remarkable performance in a wide range of NLP tasks, including information retrieval. Researchers have investigated techniques to fine-tune pre-trained Transformer models for specific retrieval tasks and integrate them with Elasticsearch for efficient document retrieval and question answering (Yang et al., 2021; Kim et al., 2022).

Scalability and Efficiency: Addressing scalability and efficiency challenges in large-scale information retrieval systems has been a key focus of recent research efforts. Studies have proposed optimization strategies for Elasticsearch deployments, such as distributed indexing and caching mechanisms, to handle the growing volume of data and user queries efficiently. Additionally, researchers have explored techniques to optimize the computational resources and model architectures of Transformer-based retrieval systems for improved scalability and performance (Zhang et al., 2021; Wang et al., 2022).

Overall, the integration of Elasticsearch with Transformer models represents a promising direction for enhancing search and information retrieval systems. By leveraging the strengths of both technologies, researchers aim to develop efficient and scalable solutions that can address the evolving needs of modern NLP applications.

# 3. FUNCTIONAL ALGORITHM AND METHODOLOGY

## 3.1 Integration of Elasticsearch with Transformer Models

To ensure a more stable and reliable integration training process for the Transformer model, there are a number of effective approaches that can be taken. First, integrate multiple Transformer models into a more powerful integrated model using model integration techniques such as Bagging, Boosting, and Stacking to reduce the risk of overfitting individual models and improve overall performance and stability. Secondly, regularization methods, such as RMSNorm normalization function and SwiGLU activation function regularization and Dropout, are introduced to control the complexity of the model and improve the generalization ability to effectively prevent the model from overfitting. In addition, appropriate loss functions, such as cross entropy loss function and mean square error loss function, are designed to measure the performance of the model and guide the training process of the model, so as to further improve the training stability and performance of the integrated model.

（1） RMSNorm normalized function

In order to make the model training process more stable, GPT-2 introduced the pre-layer normalization method compared with GPT. The first layer normalization was moved before the multi-head self-attention layer, the second layer normalization was also moved before the fully connected layer, and the position of residual connection was also adjusted after the multi-head self-attention layer and the fully connected layer. The RMSNorm normalization function is also used in the layer normalization. The formula for calculating the input vector aRMSNorm function is as follows:

$$\text{RMS}(a) = \sqrt{\frac{1}{n}\sum_{i=1}^{n} a_i^2} \quad (1)$$

$$\bar{a}_i = \frac{a_i}{RMS(a)} \quad (2)$$

In addition, RMSNorm can also introduce learnable scaling factor gi and offset parameter bi, resulting in:

$$\bar{a}_i = \frac{a_i}{RMS(a)} g_i + b_i \quad (3)$$

RMSNorm code implementation in HuggingFace Transformer library is as follows:

```
ss LlamaRMSNorm(nn.Module):    def __init__(self, hidden_size, eps=1e-6):         """
LlamaRMSNorm is equivalent to T5LayerNorm         """            super().__init__()
self.weight = nn.Parameter(torch.ones(hidden_size))
```

Modern converter-based LLMS, such as ChatGPT and Llama 2, typically go through the following three key step training process:

## 3.2 methodology

In this experiment, the to enhance sentiment text analysis using Transformer models integrated with large language models. The rapid advancement of Transformer models, such as BERT and GPT, has shown remarkable performance in various natural language processing tasks, including sentiment analysis. By leveraging the powerful capabilities of Transformer models and integrating them with large language models, we seek to improve the accuracy and efficiency of sentiment text analysis in cloud-based environments.

### 3.3.1 Create a subdataset

For a more accurate rating prediction, this article will only use enterprise data in the restaurant category. business ids for 52,268 restaurants were obtained from Business.json and 4,724,471 reviews were obtained from review.json. Some data samples are shown in Table 1. At the same time, the number of comments is not equal for each rating category, and the distribution of these data is shown in Figure 3 and Table 4. It can be found that the data is biased towards both poles, with 67.94% of reviews giving 4 and 5 stars. This unbalanced distribution has been verified in...

Table 1. Example of Yelp Dataset

| text | stars |
|---|---|
| If you decide to eat here, just be aware it is... | 3 |
| Family diner. Had the buffet. Eclectic assortm... | 3 |
| Wow! Yummy, different, delicious. Our favo... | 5 |
| Cute interior and owner (?) gave us tour of up... | 4 |
| I am a long term frequent customer of this est... | 1 |

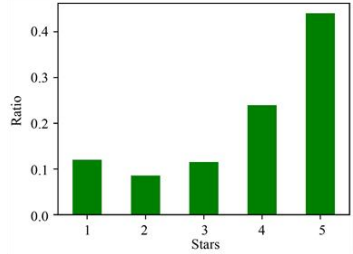

Figure 2. Rating Distribution of Yelp Dataset

According to the data in Table 1 and Figure 2, in order to process the unbalanced raw data, a balanced training data set needs to be reconstructed, and the training set, validation set and test set are divided in a ratio of 70:15:15. Thus, 1 million comments were resampled from the raw data of a training dataset, with 200,000 comments (1 to 5 stars) per category. In addition, both the validation and test sets had 200,000 comments, but they roughly followed the uneven distribution of the original data.

### 3.3.2 Input matrix model

In order for the data to enter the model correctly, the text document is first transformed into numerical data (a matrix) to build the features. Some detailed text preprocessing pipelines are given in [1], this paper uses scikit-learn [11], and the specific operations are as follows:

Count Vectorizer can convert text into a token counting matrix, while Tf-Idf Vectorizer will use Tf-Idf instead of token counting. Tf-idf is the product of the term frequency $Tf(t,y)$ and the inverse document frequency $Fyc(t)$ :

$$Tf\text{-}idf(t,d)=Tf(t,d)\times df(t) \quad (4)$$

Where $Tf(t)$ is the frequency of occurrence of the term t in document d, and $idf(t)$ is defined as:

$$idf(t) = \log\frac{n}{1+df(t)} \quad (5)$$

Where n is the total number of documents in this dataset and $df(t)$ is the number of documents in the dataset containing the term t. For binary versions of the vector, all non-zero counts are set to 1. Sparse matrices are also output from these vectors.

### 3.3 Experiment and result

3.3.1 Evaluation indicators

This article will use the following four evaluation metrics:

1) Accuracy. Represents the proportion of samples with positive prediction in the total number of samples:

$$\text{Accuracy} = \frac{TP+TN}{TP+FP+FN+TN} \quad (6)$$

2) Accuracy rate. Represents the proportion of the predicted positive samples that are actually positive in the predicted positive samples:

$$\text{Precision} = \frac{TP}{TP+FP} \quad (7)$$

3) Weighted F1 value:

$$\frac{1}{\sum_{q \in Q} |\hat{y}_q|} \sum_{q \in Q} |\hat{y}_q| F_1(y_q, \hat{y}_q) \quad (8)$$

Where Q is the set of labels, $yq$ is the subset with the q label, $y\ par\ q$ is the set predicted to be q, and

$$F_1 = \frac{2 \times Precision \times Recall}{Precision + Recall} \quad (9)$$

4) Confusion matrix. The evaluation effect is presented in the form of a matrix, denoted as C, where rows represent the true classification labels of the data, columns represent the predicted labels of the data, and $Cij$ is the proportion of the number of actual labels i but predicted j in all actual labels i.

### 3.4 Deep Bidirectional Transformers (BERT)

BERT is a language representation model that is pre-trained on the left and right context of unlabeled text for language models and next sentence prediction tasks. Such a model, when pre-trained on a large corpus of natural language texts, can be used for a wide range of training tasks after being fine-tuned with only an additional output layer. There are currently two BERT architectures available: A smaller base BERT model with 110M parameters (12 transformer blocks, 768 hidden units, 12 self-attention heads) and a larger BERT model with 340M parameters (24 transformer blocks, 1024 hidden units, 16 self-attention heads). This paper will conduct experiments on both BERT models. The training results are shown in Table 4, and the corresponding confusion matrix is shown in Figure 4 . In terms of setup, the maximum sequence length is 128 and the batch size is 16.

It can be found that BERT (base, large) is 1.02% more accurate than BERT (base, cased) and the weighted F1 value is improved by 0.0015 cased. The former has more parameters than the latter, but has twice the running time.

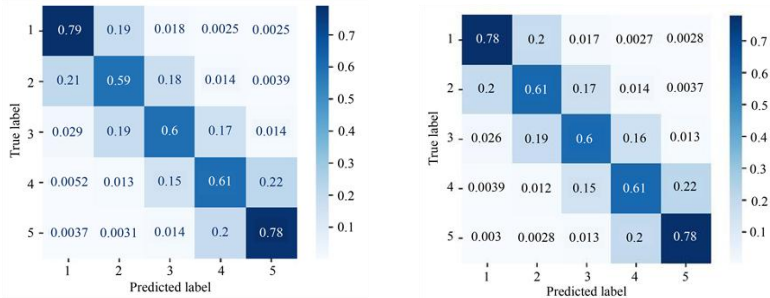

Figure 4 . Confusion Matrix of Cased Large BERT on Test Set

According to the above results, the DistilBERT model is a distillation of BERT model, in which 97% of BERT language understanding ability is retained and 60% of performance is improved. The results on the test set are shown in Table 4, and the confusion matrix is shown in Figure 4. It can be found that the accuracy of the DistilBERT model is reduced by about 1.36% compared to the multi-parameter BERT model, but it is twice as fast as the basic BERT model. When computing resources are limited, DistilBERT will be an attractive option.

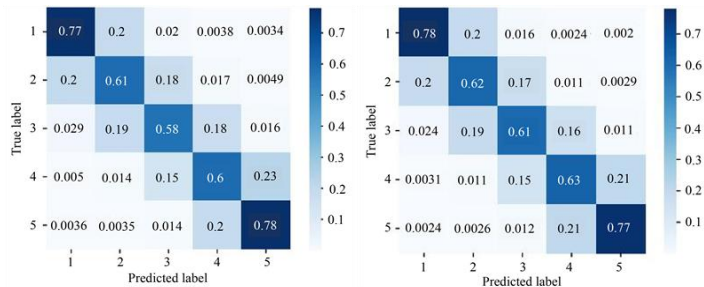

Figure 4 . Confusion Matrix of RoBERTa on Test Set

This experiment will carry out pre-training procedures and has achieved advanced results on several NLP tasks. In this paper, pre-trained roberta-base architecture was used for experiments. The relevant experimental results are shown in

Figure 5, and the confusion matrix is shown in Figure 5. It can be found that this model is optimal. Compared with the multi-parameter BERT model, the accuracy rate is increased by 0.62% and the score is higher when the two models have similar running time. This suggests that the RoBERTa model would be preferable for higher predictors when computing resources are not considered.

### 3.5 Experimental conclusion

This study analyzes Yelp's dataset for rating predictions, employing classical machine learning models and advanced Transformer-based models like BERT, DistilBERT, and RoBERTa for emotion prediction. Results indicate that DistilBERT exhibits advantages in training time, while multi-parameter BERT and RoBERTa models demonstrate improved classification accuracy, reaching 69.75%. The study underscores the effectiveness of Transformer pre-training models in emotional text processing, paving the way for future multimodal model exploration, albeit with room for improvement due to computational constraints.

## 4. CONCLUSION

In this paper, we delve into the application prospects and challenges of large language models (LLMs) in natural language processing. First, we introduced the fundamentals of LLMs and the key role of the Transformer model, highlighting their importance in tasks such as translating, summarizing, and generating languages. We then analyzed the challenges LLMs faces in cloud environments, including scalability and efficiency issues, and proposed key technologies such as Elasticsearch and Transformer models as solutions. Through the experimental verification of Yelp data set, we find that the machine learning method and Transformer model have achieved remarkable results in emotional text processing, which provides an important reference for future research and practice.

In the future, we will continue to explore the following directions: First, we will work to further optimize the performance of the model, including improving the classification effect and reducing the training time. Secondly, based on the Transformer pre-training model, we will further explore the application of multi-modal models to improve the efficiency and accuracy of language processing. Finally, we will also pay attention to the performance of LLMs in practical applications, and plan to apply these technologies to a wider range of fields to bring more practical value to society. In general, this paper provides important ideas and directions for the research and application of large-scale language models in the field of natural language processing, and contributes to the development of artificial intelligence technology.